\documentclass[11pt,twoside]{article}
\usepackage{./asp2010}

\usepackage{epsfig}
\resetcounters

\begin{document}

\title{A comparison of oscillation frequencies of rotating stars obtained by one- and two-dimensional calculations}
\author{Hideyuki Saio$^{1,2}$ and Robert G. Deupree$^2$
\affil{$^1$Astronomical Institute, Graduate School of Science, Tohoku University, Sendai, Japan}
\affil{$^2$Institute of Computational Astrophysics, Saint Mary's University, Canada}
}

\begin{abstract}
We discuss how the p- and g-mode frequencies calculated for rotating stars are affected by the way of  including the centrifugal deformations.
We find that p-mode frequencies are sensitive to the accuracy in treating the centrifugal deformation, while the effect becomes smaller for smaller frequencies in the g-mode range.
\end{abstract}

\section{Oscillations of rotating stars}
For a static spherical star, the angular dependence of the eigenfunction of an oscillation mode is represented by a single spherical harmonic $Y_\ell^m(\theta,\phi)$, and the frequencies depend on the radial order $n$ and $\ell$ but independent of $m$; i.e. degenerate with respect to azimuthal degree $m$.  
For a rotating star, the effects of Coriolis force and centrifugal deformation of the equilibrium structure enter into the equations for stellar oscillations.
The momentum equation for linear oscillations in the co-rotating frame of a uniformly rotating star with an angular frequency $\Omega$ may be written as;
$$-\omega^2\vec{\xi} + 2i\omega\vec{\Omega}\times\vec{\xi} =
-{1\over\rho_0}\nabla P' + {\rho'\over\rho_0^2}\nabla P_0 -\nabla\psi', \eqno(1)$$
where $\psi$ is the gravitational potential, $(')$ means Eulerian perturbation, and $\omega$ is the angular frequency of oscillation in the co-rotating frame.
The second term of the left-hand-side of equation (1) comes from Coriolis force. 
This lifts the degeneracy of frequencies with respect to  $m$, and makes it impossible to represent the angular dependence of the eigenfunction of an oscillation mode by a single spherical harmonic.
The rotational deformation makes the latitudinal gradients of equilibrium quantities such as $\partial P_0/\partial\theta$ nonzero.  
In addition, due to rotational advection (or coordinate transformation) observational (or inertial) frequency $\sigma$ differs from frequency $\omega$ in the co-rotating frame as
$$\sigma = \omega - m\Omega. \eqno(2)$$
In this paper we adopt the convention that negative $m$ corresponds a prograde mode.

\begin{figure}[t]
\begin{center}
\epsfig{file=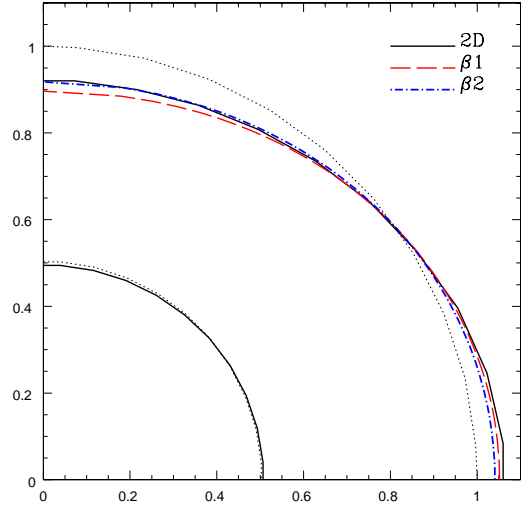,
width=0.6\textwidth}
\end{center}
\caption{Isobaric surfaces of a 2D $10M_\odot$ ZAMS model rotating at an equatorial velocity of 360~km/s (solid lines) are compared with those from approximations '$\beta1$' and '$\beta2$'.
The rotation speed corresponds to $\Omega = 0.49\sqrt{GM/R^3}$, where $R$ is the mean radius. 
Dotted lines indicate spheres.}
\label{deform}
\end{figure}

Although using two dimensional (2D) models \citep{ree06,lov08} is the way to obtain accurate frequencies, it is time-consuming and nonadiabatic analysis is not available yet.
Therefore, one dimensional analyses where deformation effects are approximately included have some merits.
We discuss in this paper the effects of the treatments of the deformation on the oscillation frequencies of rotating stars.
We have adopted the followingthree ways; 
\begin{itemize}
\item 2D: Latitudinal gradients are obtained directly from 2D equilibrium models \citep{cle98,lov08}
\item Assuming the structure is barotropic, the distance from the center of an isobar surface is written as $r(p)=a(p)[1+\beta P_2(\cos\theta)]$.
\begin{itemize}
\item $\beta1$:  Obtain $a(p)$ and $\beta(a)$ by fitting with 2D models.
\item $\beta2$:  Obtain $a(p)$ from spherical models calculated with mean centrifugal forces, and $\beta(a)$ from Chandrasekhar expansion \citep{lee95}   
\end{itemize}
\end{itemize}
We note that 2D and $\beta1$ are based on the same 2D models, while $\beta2$ based on spherical models.

We express the spatial dependence of the displacement vector $\vec{\xi}$ and a scaler variable $f'$ as
$$\vec{\xi}=\sum_{j=1}^J[S^{j}Y_{l_j}^m +H^{j}\nabla_{\rm h}Y_{l_j}^m + T^j(\nabla_{\rm h}Y_{l'_j}^m)\times \vec{e_r}], \qquad f' = \sum_{j=1}^J f'^{j}Y_{l_j}^m,$$
where
$$\nabla_{\rm h}={\partial\over\partial\theta}+{1\over\sin\theta}{\partial\over\partial\phi},
\qquad l_j = |m| + 2(j-1) + I,  \quad l'_j = l_j +1-2I,
$$
with $I=0$ for even modes (in which scaler variables are symmetric with respect to the equator) and $I=1$ for odd modes.
The terms proportional to $T^j$ represent a toroidal displacement which is needed because of the Coriolis force term in equation (1). 
Obviously a single latitudinal degree $\ell$ of $Y_\ell^m$ is not a good parameter anymore for a rotating star.
We designate the type of the latitudinal dependence of a mode by using $\ell_0$ which corresponds to the latitudinal degree at $\Omega=0$. 
We compare the frequencies obtained for $10M_\odot$ zero-age main sequence models with various rotation speeds.

\section{Results}

\begin{figure}[t]
\begin{center}
\epsfig{file=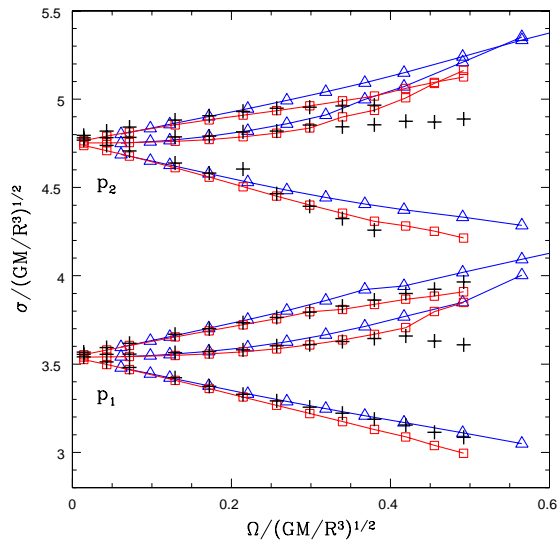,
width=0.6\textwidth}
\end{center}
\caption{Frequencies in the inertial frame of nonradial p$_1$ and p$_2$ modes of $\ell_0 = 1$ as a function of rotation frequency $\Omega$ obtained by different methods. 
Each mode have three frequencies corresponding to $m=-1, 0, 1$ ($m=-1$ is the highest, see eq.(2))
Plus signs are for 2D calculations, squares and triangles are from the methods of $\beta1$ and $\beta2$ (see text). 
}
\label{pmodes}
\end{figure}

Figure~\ref{pmodes} compares frequencies in the inertial frame for p$_1$ and p$_2$ modes of $\ell_0=1$ at various $\Omega$ obtained by different methods discussed in the previous section.
Deviations in frequencies between b2 (triangles) and 2D models (plus) become appreciable at $\Omega\approx 0.2\sqrt{GM/R^3}$, which corresponds to a equatorial velocity of $\sim150$km/s.
Among the frequencies for different azimuthal order $m$, deviations for axisymmetric ($m=0$) modes are largest, while deviations for sectoral ($m=\pm1$) are remained small.

The deviations between b1 (squares) and 2D sequences become appreciable around $\Omega\approx 0.35\sqrt{GM/R^3}$ for p$_2$ modes and  $\sim0.4\sqrt{GM/R^3}$ for p$_1$ mode.
This figure clearly indicates that to obtain accurate frequencies of p-modes (higher modes in particular) of rapidly rotating stars it is necessary to accurately include centrifugal deformations.

\begin{figure}
\begin{center}
\epsfig{file=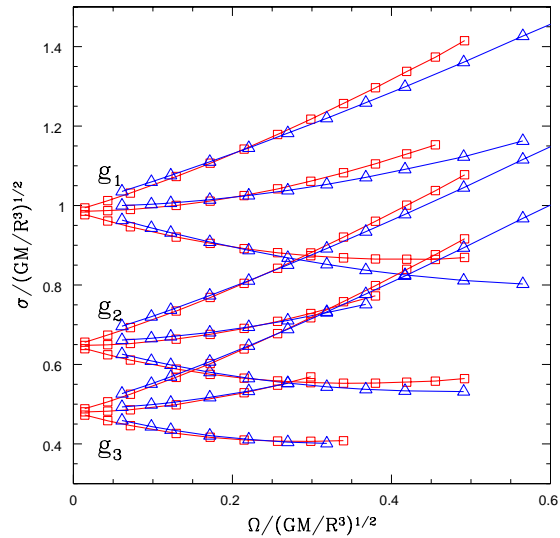,
width=0.6\textwidth}
\end{center}
\caption{The same as Fig.~\ref{pmodes} but for g$_1$, g$_2$, and g$_3$ modes of $\ell_0=1$.
}
\label{gmodes}
\end{figure}

Figure~\ref{gmodes} compares frequencies of low-order g-modes of $\ell_0=1$ obtained using the assumptions $\beta1$ (squares) and $\beta2$ (triangles).
For g$_1$ modes differences are appreciable for $\Omega \gtrsim 0.3\sqrt{GM/R^3}$.
For g$_2$ and g$_3$ modes, on the other hand, the differences remain small even for $\Omega$ as large as $\sim 0.4\sqrt{GM/R^3}$, which corresponds to an equatorial velocity of $\sim300$km/s.
This indicates that for high-order g-modes, which are confined largely in the core, rotational deformation effect is not important. 
Instead,  they are affected strongly by Coriolis force whose effect increases as a function of $\Omega/\omega$  

\section{Conclusion}
We have examined the effects of the treatments of centrifugal deformation on the frequencies of nonradial oscillations of rotating stars.
It is found that p-mode frequencies are sensitive to the treatments of the deformation, while g-mode frequencies are insensitive.
The sensitivity difference can be understood by the fact that p-modes are confined in the outer envelope where the rotational deformation is significant, while g-modes are confined in the core where the rotational deformation is small.


\bibliography{pref}

\end{document}